\documentclass[a4paper,10pt]{spie}
\usepackage{graphicx}
\usepackage{amsmath}
\usepackage{amssymb}
\usepackage{textcomp}

\title{Expected performance of a Laue lens based on bent crystals }

\author{Vineeth Valsan\supit{a,b}, Filippo Frontera\supit{a}, Enrico Virgilli\supit{a}, Vincenzo Liccardo\supit{a,b}
\skiplinehalf
\supit{a} \textit{Physics Department, University of Ferrara, Via Saragat 1, Ferrara, Italy}; \\
\supit{b} \textit{Universit\'e de Nice Sophia Antipolis, Nice, Cedex 2, Grand Chateau Parc Valrose, France}.
}

\authorinfo{Contact information:E-mail: valsan@fe.infn.it}

\begin{document}

\maketitle

%-------------------------------------

\begin{abstract}
In the context of the LAUE project devoted to build a long focal length focusing optics for soft gamma-ray astronomy (70/100 keV to $>$600 keV), 
we present results of simulation of a Laue lens, based on bent crystals in different assembling configurations 
(quasi-mosaic and reflection-like geometries). 
The main aim is to significantly overcome the sensitivity limits of the current generation of gamma-ray telescopes and improve the imaging capability.
\end{abstract}


%-------------------------------------

\section{The Laue Project}

The LAUE project, supported by ASI, the Italian Space Agency, is devoted to the building of a 
wide range (70/100 keV to $>$600 keV) focusing space telescope with a focal range of 100 m.
The lens assembling technology is very much improved thanks to the experience gained from the previous project HAXTEL\cite{Frontera08} 
with two prototypes of the lens\cite{Frontera10,Virgilli11}. 

Various crystals has been tested for the lens using the LARge Italian X-ray (LARIX) Facility of the physics Department of University of Ferrara.

\section{Reflectivity of bent crystals: Theory}

The diffraction profile of a bent crystal is different from that of the perfect crystal. When a crystal is bent, 
the standard dynamical theory related to the perfect crystal cannot be applied. There are many ways of curving 
a crystal. Most the methods used have been described in [\citenum{Buffagni12}, \citenum{Guidi11} and \citenum{Marchini11}].
Curving a crystal changes the angular distribution of the incoming beam at the entrance surface and
it also deforms the regular atomic spacing inside the crystal. 
The reflectivity of such bent crystals can be simulated by using many methods, but each method has its own limitations.
Some of the methods are described below.

\subsection{Multi Lamellar Method}

In this method, the crystal is supposed to be consisting of several layers of thin perfect crystal tiles.
Each of these layers are assumed to be misaligned in such a way that they follow the curvature of the entire crystal.
The dynamical theory of plane crystals, as described by Zachariasen[\citenum{Zachariasen}], can be used to calculate the
reflected and transmitted rays because each layer is assumed to be perfect and thin. 

A detailed description of the method used for calculating the crystal reflectivity was given by Boeuf et al.~[\citenum{Boeuf78}] 
in the case of X-ray diffraction in reflection configuration, and by Sanchez del Rio [\citenum{Sanchez98}] 
in reflection and transmission configuration of the crystals.

\subsection{Takagi-Taupin(TT) Equations}

Using TT equations \cite{Takagi69, Taupin64}, the behavior of the electromagnetic field inside the crystal is described.
Following Sanchez del Rio et.~al.[\citenum{Sanchez98}], these equations are given by: 

\begin{equation}
 \frac{\partial}{\partial s_0}D_0(\vec{r}) = -i\pi k P \varPsi_H D_h(\vec{r})
\end{equation}

\begin{equation}
 \frac{\partial}{\partial s_h}D_h(\vec{r}) = -i\pi k P \varPsi_H D_0(\vec{r}) + 
2\pi i \left\lbrace k \beta_h -\frac{\partial}{\partial s_h} \left( \vec{h} . \vec{u}(\vec{r}) \right) \right\rbrace 
\end{equation}

where $D_0$ and $D_h$ respectively are the amplitudes of the transmitted and diffracted fields, 
$k$ is the wavevector, $\vec{h}$ is the reciprocal lattice vector, $\vec{u}$ is the displacement vector of the deformation,
and $\beta_h $ $=$ $\dfrac{ |\vec{k}_h |^2 - |\vec{k}_0 |^2 }{ 2 \vec{k}^2 }$, $\vec{k}_h$ and $\vec{k}_0$ are the
diffracted and transmitted wavevectors respectively.

The solution of these equations gives the diffracted and transmitted beams of a distorted crystal. 
Different methods have been adopted to solve these equations for the Laue as well as for the Bragg case [\citenum{Boeuf78}].
Sanchez del Rio et. al.[\citenum{Sanchez98}] has given the numerical solution of these equations using Finite Element Method(FEM).

\subsection{Penning-Polder(PP) Method}

In this method, put forth by Penning and Polder [\citenum{Penning61}], the complete bent crystal is assumed to be made of many perfect crystal-parts. 
Then the dynamical theory of undistorted crystal is applied to all these perfect crystal-parts. 
The X-ray beam is assumed to be propagating through the distorted crystal as a pseudo-plane block wave,
i.e. PP method exploits the wave nature of X--rays inside the crystal. This theory also assumes that, while passing from one 
part of the crystal to the next, the wavefield is preserved.  
This method is only applicable to the Laue case due to the anomalous absorption effect on the total reflection angle. 

Following the paper by Sanchez Del Rio et. al [\citenum{Sanchez98}], the reflectivity of a symmetric crystal is given by:

% the constant strain gradient follows the relation
% 
% \begin{equation}
%  \left[ \xi_i + \frac{1}{\xi_i} + \frac{2\beta x}{\tan \theta_b} \right]^2 - \left[ 2 \beta z + \xi_i - \frac{1}{\xi_i} \right]^2 = 4
% \end{equation}
% 
% And the reflectivity of a symmetric crystal is given by:

\begin{equation}
 R = \frac{\xi_{e_j}^2}{\xi_{i_j}^2 + b} \frac{b}{\xi_{i_j}^2 + b} \exp{ \left\lbrace -\frac{\mu t}{\cos(\theta_b - \pi/2)} 
\left[  1 + \frac{b-1}{2t\beta}(\xi_{e_j} - \xi_{i_j}) + b\frac{P\epsilon}{\beta t} \ln\frac{\xi_{e_j}}{\xi_{i_j}}  \right] \right\rbrace }
\end{equation}

Where $\xi_i$ represents the ratio of amplitudes of the transmitted and diffracted plane wave components, the subscripts $i$ and $e$ represents
the entrance and exit surface respectively. For an elastically isotropic crystal, the strain gradient is given by:

\begin{equation}
 \beta = \frac{b - 1}{P_\rho \varPsi_H} \left[ 1 + (\cos\theta_b -1)\frac{1+\nu}{2} \right]
\end{equation}

Where all the parameters has the same meaning as given by [\citenum{Sanchez98}]. 

\subsection{Malgrange's extension of PP theory}

The Malgrange treatment is an extension of the Penning-Polder method.
It is valid for bent crystals having large and homogenous curvature radius.
In this theory the strain gradient $\beta$ describes the distortion of the diffraction planes and is given by:

\begin{equation}
 \beta = \frac{\Omega}{T_o (\delta/2)}
\end{equation}

where $\Omega$ is the total bending angle. It corresponds to the FWHM of the angular distortion distribution of the lattice planes, 
in case of quasi-mosaic crystals. $T_o$ is the thickness of the crystals 
and $\delta$ is the Darwin's width. 

Above a critical value, $\beta_c$, of the strain gradient given by $\beta_c = \dfrac{\pi}{2\Lambda_0}$, where $\Lambda_0$ is the extinction length, 
the intensity of the diffracted wave decreases because of the creation of a new wavefield.
For a uniform curvature  $C_p$ of the lattice planes across the crystal thickness 
and with the condition that the value of the strain gradient $\beta$ 
is larger than the critical value $\beta_c$,
the peak reflectivity, $R_{peak}$ of the curved crystal is given by: 

\begin{equation}
 R_{peak} = \left[ 1 - \exp{\left( -\dfrac{\pi^2 d_{hkl}}{C_p \Lambda_0^2} \right)}  \right] 
\left[\exp\left( -\dfrac{\mu \Omega}{C_p \cos\theta_b}\right) \right]
\end{equation}

where $C_p = \dfrac{\Omega}{T_o}$,  $d_{hkl}$ is the spacing of the lattice planes ($hkl$), $\mu$ is the absorption coefficient at a given energy
and $\theta_b$ is the Bragg angle.
This equation has been used to simulate the performance of the Laue lens petal prototype using bent crystals in this paper.

\section{Material selection and crystal geometry}

Bent crystals of Ge(111) and GaAs(111) have been selected for the LAUE project, both in transmission configuration. One of them (GaAs) 
has a mosaic structure with about 20 arcsec mosaicity, while the other (Ge) is a perfect crystal that has been bent. The bending technology  adopted for Ge is the surface grooving \cite{Guidi11}, while that adopted for GaAs is a lapping process \cite{Marchini11,Buffagni12}.
The selected tiles will be provided respectively by the ``Laboratorio Sensori e Semiconduttori'' (LSS) 
of Ferrara and by the ``Istituto Materiali per Elettronica e Magnetismo'' (IMEM) of Parma (Italy). 
A description of the material's properties is given in the Table \ref{tab:material}.

We have investigated the expected reflectivity of these two families of crystals. While GaAs has a mosaic structure, Ge(111)
shows a quasi-mosaic structure (see Ref.~\citenum{Liccardo12}).

\begin{table}[h]
 \centering
  \begin{tabular}{|l|c|c|}
   \hline
   Properties    		&  GaAs  		&  Ge  				\\ \hline  \hline
   Chosen lattice planes  		&    (111)    		&   (111)   			\\ \hline
   Crystal structure  		&    Mosaic, Curved    	&    Quasi-mosaic, Curved  	\\ \hline
   Cross-section (mm$^2$) 	&  $30\times 10$   	&   $30\times 10$  		\\ \hline
   Thickness (mm)        		&    2    		&   2   			\\ \hline
   Mosaicity FWHM (arcsec) 		&    20    		&   4   			\\ \hline
   External Curvature Radius (m) 	&   40    		&   40   			\\ \hline
  \end{tabular}
 \hspace{10cm}
 \caption{Properties of the selected materials to be used as elements for the LAUE project.}
 \label{tab:material}
\end{table}

\begin{figure}[ht]
  \begin{minipage}[b]{0.47\linewidth}
    \centering
    \includegraphics[width=0.7\textwidth]{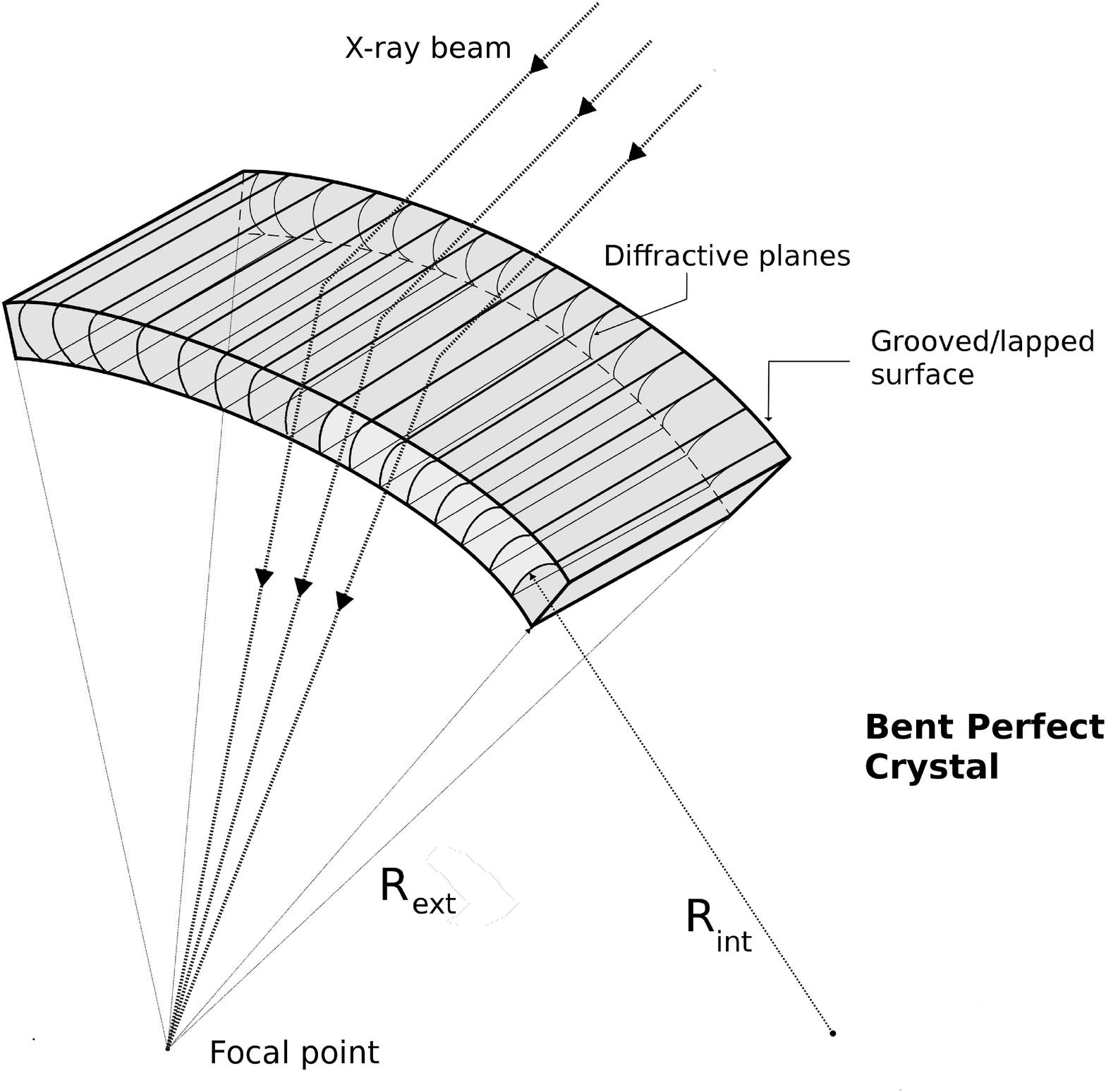}
    \caption{Quasi-mosaic crystal principle. The FWHM in this case depends upon the internal curvature of the atomic 
planes caused by the bending of the crystal.}
    \label{fig:quasimosaic}
  \end{minipage}
%
% Figure 2
%  
\hspace{0.5cm}
  \begin{minipage}[b]{0.47\linewidth}
    \centering
    \includegraphics[width=0.7\textwidth]{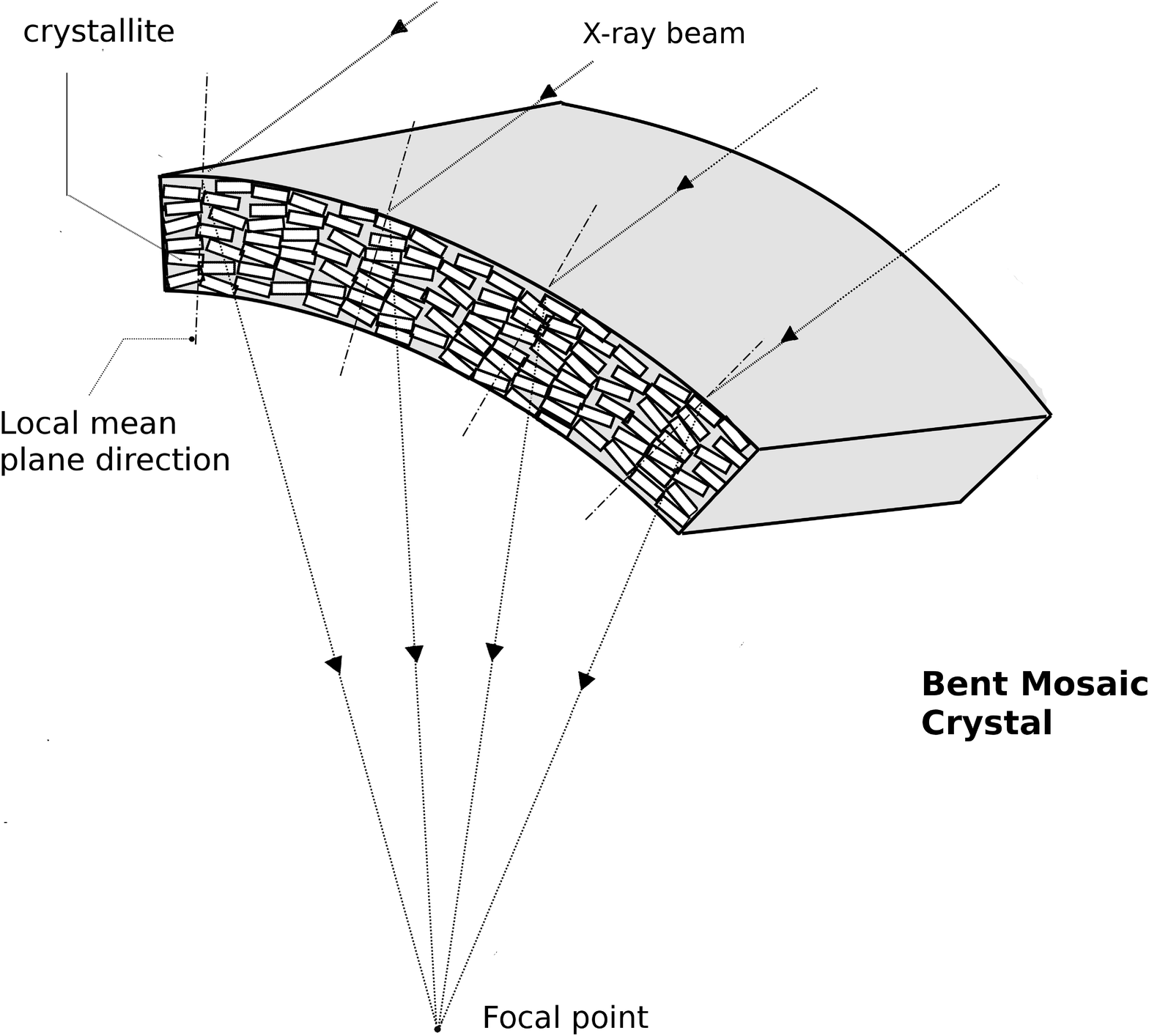}
    \caption{Bent crystal principle in the Mosaic configuration. 
The FWHM of the reflectivity of these crystals depends upon the mosaicity of the crystals.}
    \label{fig:bentmosaic}
  \end{minipage}
\end{figure}

Indeed, from the dynamical 
theory of diffraction of bent crystals, it results that the bending of a perfect crystal creates a curvature of the lattice planes (111) of Ge, 
with a ratio between the internal curvature and external curvature radii of 2.6. This gives rise to a quasi-mosaic configuration of the bent crystal 
(see Fig.~\ref{fig:quasimosaic}). This effect is not valid for all lattice planes. For example, if the diffracting planes in transmission configuration are (220), 
the quasi-mosaicity is created.

The advantage of bent perfect crystals is that their reflectivity  exceeds the theoretical limit of 50\%, valid with for flat perfect crystals. 

Concerning mosaic crystals like GaAs, the reflectivity limit of 50\% is still conserved, even if their are bent. The advantage of bending is its focusing capability. When a mosaic crystal is bent, the mean crystallite Gaussian distribution is expected to continuously change along 
the crystal (Fig. \ref{fig:bentmosaic}). Experimental tests confirm the expectations \cite{Liccardo12}.

%
% Figure 1
%

\subsection{Crystal thickness}

Simulations based on crystal geometry were performed using the Malgrange equations to define the best thickness 
for each specimen. According to Malgrange's theory, in a flat crystal the reflectivity value is related to the extinction 
length $\Lambda_0$ (e.g., Ref.~\citenum{Frontera10}) that depends on the diffracted energy and crystal material. 

\begin{figure}[ht]
  \begin{minipage}[b]{0.48\linewidth}
    \centering
    \includegraphics[width = 0.8\textwidth]{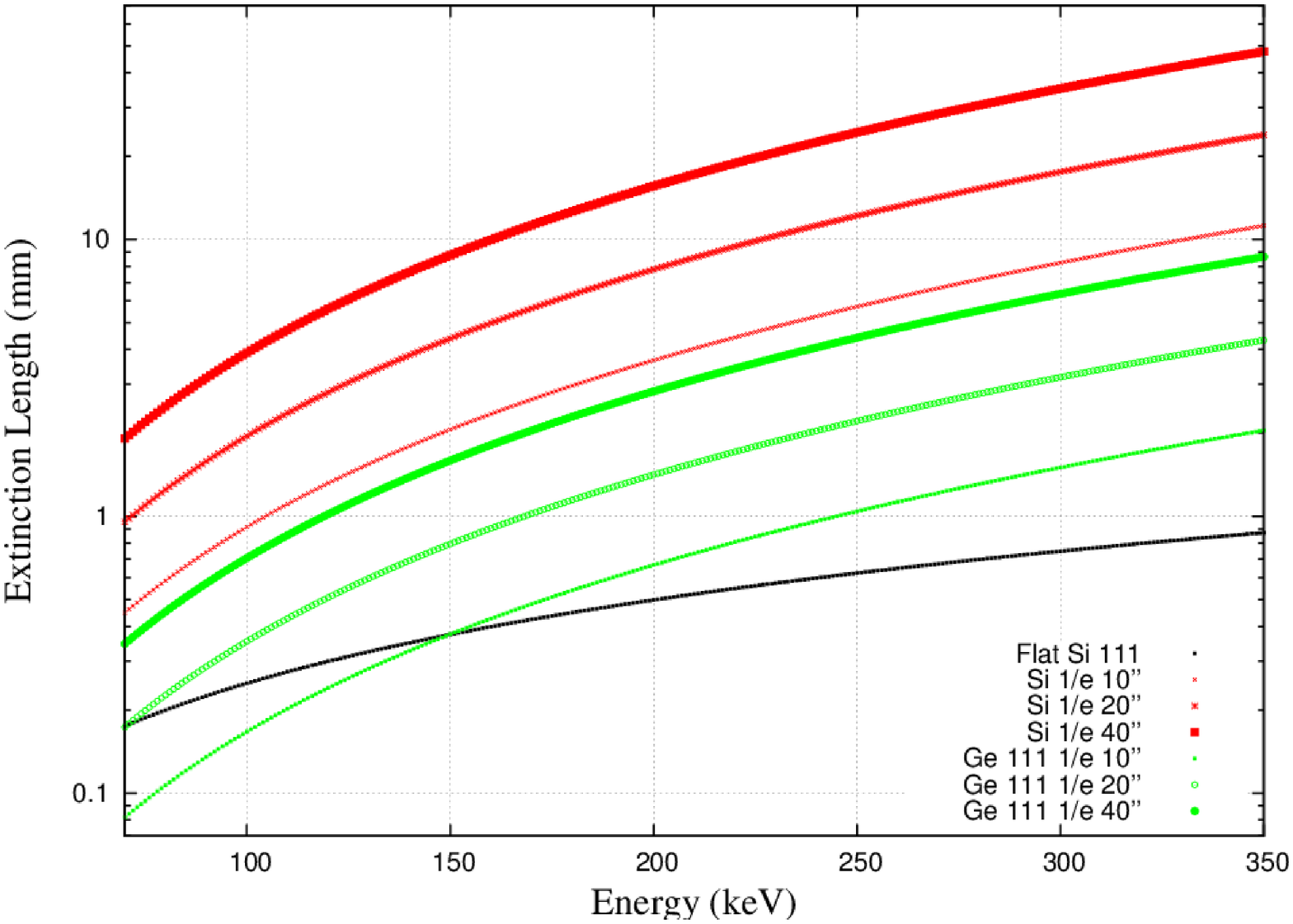}
    \caption{Extintion length vs. energy for bent Silicon (red line) and Germanium (green) compared with that of flat Silicon (black)}
    \label{fig:extlngth}
  \end{minipage}
%
% Figure 4
%  
\hspace{0.5cm}
  \begin{minipage}[b]{0.48\linewidth}
    \centering
    \includegraphics[width=0.8\textwidth]{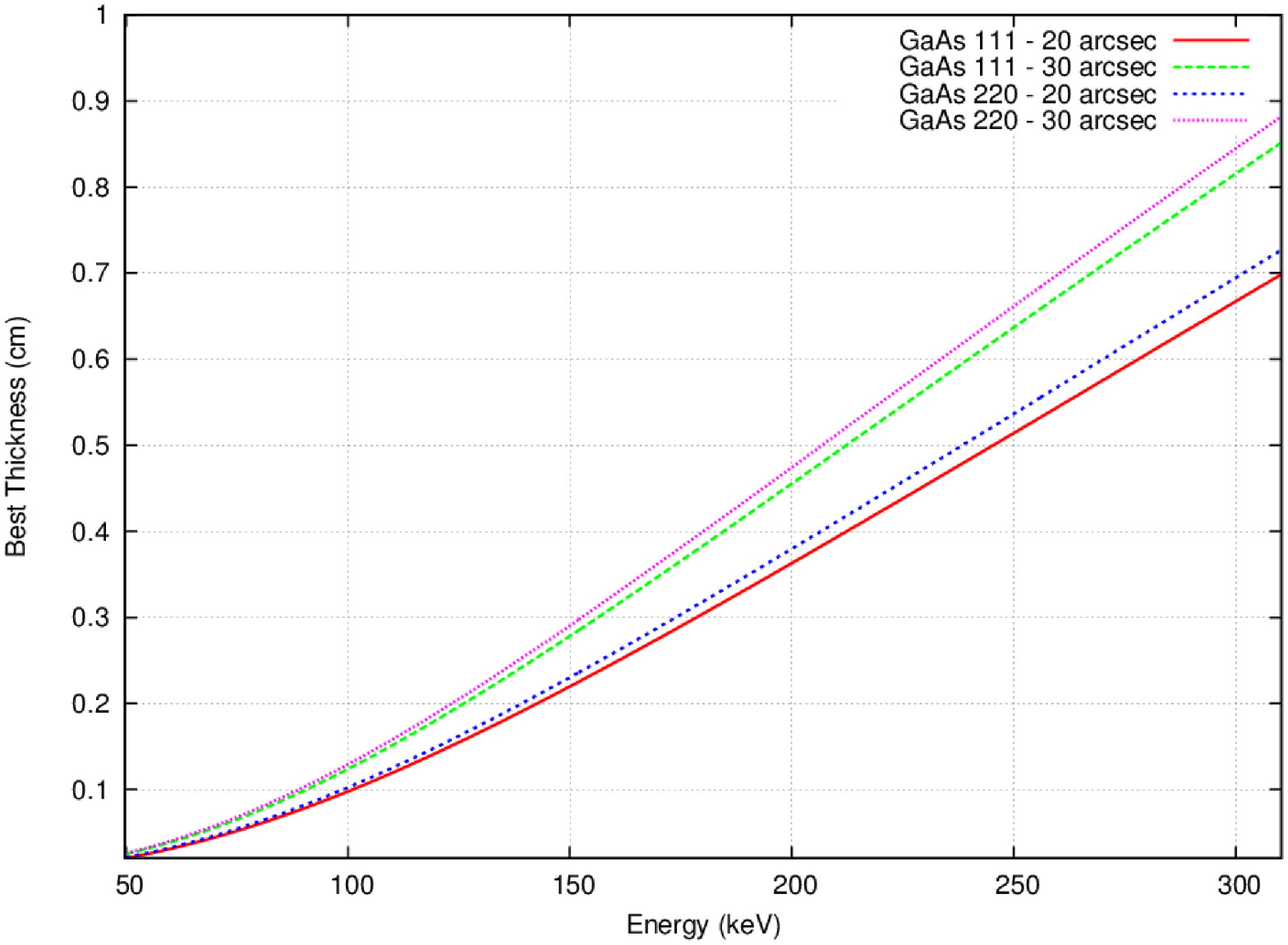}
    \caption{Best thickness as function of the energy for crystals made of Gallium Arsenide with different angular spread (20 and 30 arcsec)}
    \label{fig:TvsE}
  \end{minipage}
\end{figure}

Using the Malgrange theory for bent crystals, the extinction length was determined for potentially suitable 
materials. In Figure \ref{fig:extlngth}, it is shown the variation of the extinction length $\Lambda_0^b$ of bent crystals as a function of the energy for Si(111) and Ge(111), for different values of the internal  
curvature. 

%
% Figure 3
%

Unfortunately bent crystals of Si(111) cannot be used for LAUE project. Indeed bent Si(111) with, e.g.,  an angular spread of 10 arcsec, shows an extinction length 
of $\sim$ 1 mm and 3.8 mm at 100 and 200 keV, respectively. Given that the  thickness requested for getting a  good  reflection efficiency requires a 
few extinction lengths, it results that the thickness requested for Si(111) ranges
from 3 to 12 mm, going from 100 keV to 200 keV, respectively. 
Unfortunately curved crystals of Si(111) with such thicknesses are not attainable due to the limitation of the current bending technology. 

The smaller extinction length of curved Ge(111) makes  the required thickness (about 2 mm) feasible.

The relation between the thickness and the energy for bent mosaic GaAs has also been reported in Figure \ref{fig:TvsE}. 
Both (111) and (220) planes of GaAs, with 20 and 30 arcsec, are compared. Tiles made of bent GaAs will be used at low energies ($<$ 150 keV), where a thickness of 2 mm is more than satisfactory, with  an angular spread of 20 arcsec.

\section{Lens Petal Prototype}

Samples of bent Si(111) with quasi-mosaic structure and of bent GaAs (111) with quasi-mosaicity, have been tested in the small 
LARIX facility of the Physics Department of University of Ferrara, with preliminary results reported in Ref.~[\citenum{Liccardo12}].

The designed Laue lens is made of 20 petal structures of equal surface area, each one 
with a subtended angle of $18$\textdegree. As a result of the LAUE project, one of these petals will be built with 
properties as reported in Table \ref{tab:petal}.

\begin{table}[h]
 \centering
  \begin{tabular}{|c|c|}
   \hline
   Energy Range   		&  96 -- 300 keV	\\ \hline  
   Focal Length 		&    20 m		\\ \hline
   Inner Radius 		&    30 cm    	 	\\ \hline
   Outer Radius			&  80 cm   	  	\\ \hline
   Total Number of Crystals     &    288  		\\ \hline
   Number of Rings		&    18   		\\ \hline
   Angle Subtended at the Center	&    18\textdegree   	\\ \hline
  \end{tabular}
 \hspace{10cm}
 \caption{Lens petal properties}
 \label{tab:petal}
\end{table}

\begin{figure}[ht]
   \centering
    \includegraphics[scale = 0.5]{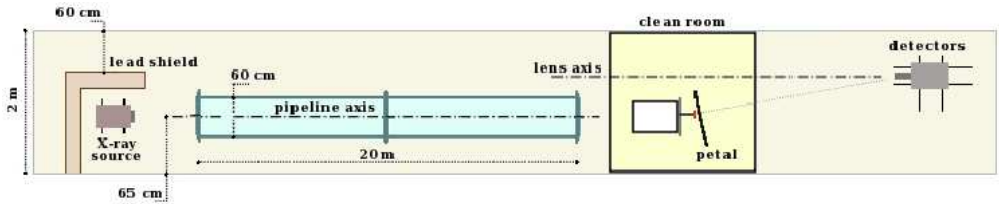}
    \caption{The larix facility where the lens petal prototype will be build.}
    \label{fig:tunnel}
\end{figure}

Crystal tiles, selected for LAUE project, are requested to show a rectangular cross section of 30 $\times$ 10 mm$^2$, with the longer dimension being set
radially along the lens radius. Depending on the current beam-line (with a diameter of 60 cm) and on the available 
X-ray source for building and testing the petal (Max. energy of 320 keV), the lens petal has been designed to focus energy 
in the range $\sim$ 100 -- 300 keV. The petal prototype will be built inside the newly developed LARIX facility installed 
at the Physics Department of the Ferrara University. A sketch of the facility is shown in figure \ref{fig:tunnel}. 
Refer to Frontera et. al. [\citenum{Frontera12}] for more detailed information about the facility.

\begin{figure}[ht]
  \begin{minipage}[b]{0.46\linewidth}
    \centering
    \includegraphics[scale = 0.3]{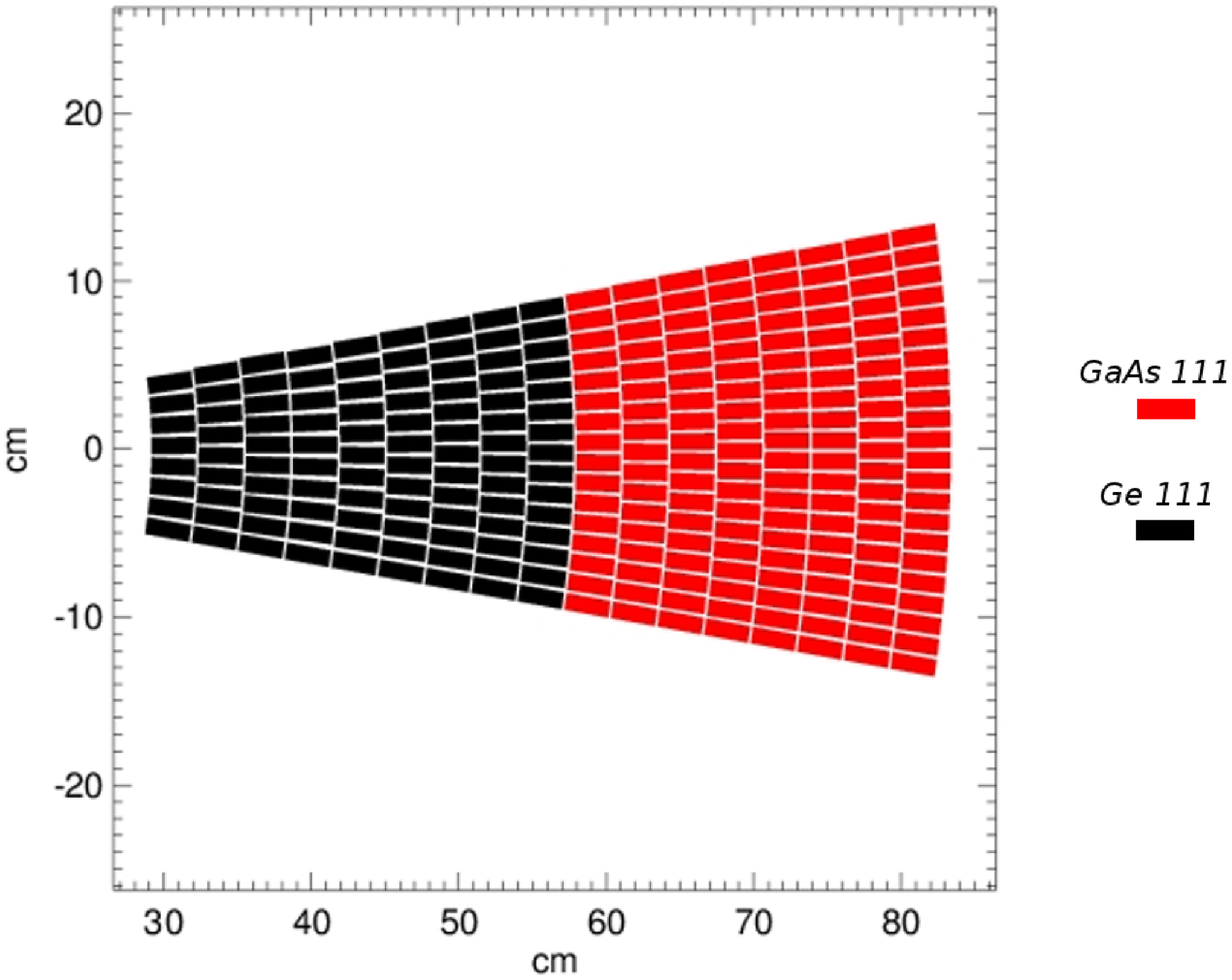}
    \caption{Illustration of the Laue lens petal made of Ge(111) and GaAs(111) with an energy pass-band of 96 -- 300 keV.}
    \label{fig:petal}
  \end{minipage}
%
% Figure 7
%  
\hspace{0.5cm}
  \begin{minipage}[b]{0.46\linewidth}
    \centering
    \includegraphics[scale = 0.3]{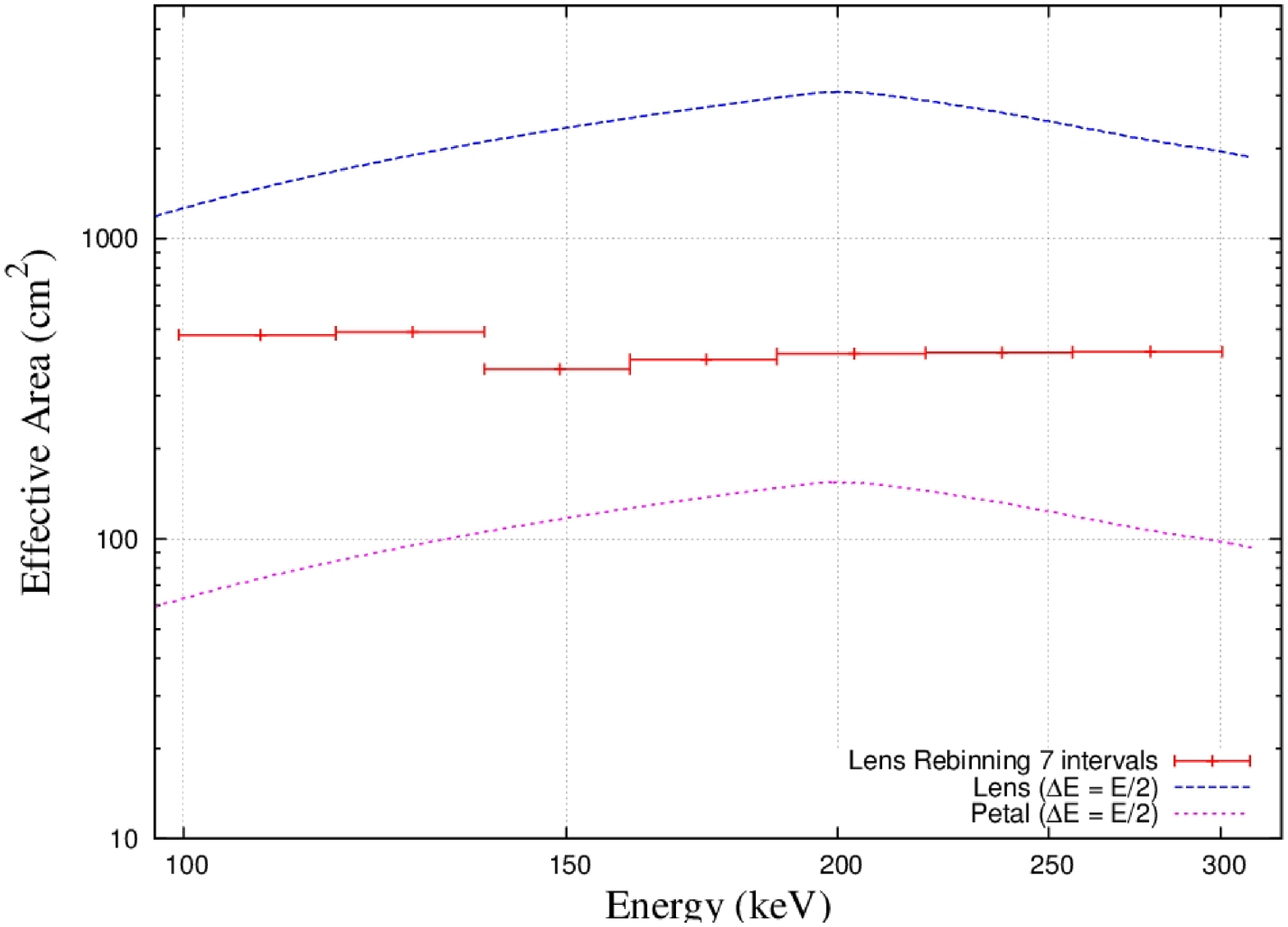}
    \caption{Effective Area for a Laue lens made of Ge(111) and GaAs(111) with an energy pass-band of 96 -- 300 keV.}
    \label{fig:ea}
  \end{minipage}
\end{figure}

As illustrated in Figure \ref{fig:petal}, the petal will be made of Ge(111) placed at lower radius (for higher energies) 
and GaAs(111) crystals at higher radius (for lower energies). In order to maximise the effective area of the lens, the
gap between the tiles was set to be less than 0.5 mm, also taking into account the process of placing the tiles side by side on 
the lens frame. Using a proper python software for simulation of the overall behaviour of both the lens petal and the 
total Laue lens, the effective area has been calculated and it is shown in Fig. \ref{fig:ea}.

%
% Figure 6 
%

\section{Conclusions}

The Laue project devoted to build Laue lenses for hard X-/soft $\gamma$-rays represents an important step forward
in the X/Gamma-ray Astronomy, being the first attempt to use bent crystals to build Laue lenses for space applications.
Focusing the radiations is so far the only possibility to increase the sensitivity and imaging capability (in terms of PSF)
with respect to the current missions. Our goal is to achieve, in the operational band of the Laue lens(80 -- 600 keV) a sensitivity of 
a few times $10^{-8}$~photons~cm$^{-2}$~s$^{-1}$~keV$^{-1}$.

In this paper, simulations were perfomed on the basis of Malgrange's expansion of 
Penning-Polder theory for curved crystals to estimate the best thickness to be used for building a Laue lens petal
focusing X-rays in the 100 -- 300 keV energy band.

The sensitivity of the petal prototype depends upon the Point Spread Function (PSF) which is
significantly improved using bent perfect and bent mosaic crystals in place of flat mosaic crystals. In addition, 
the lens sensitivity increases proportionally to the effective area that has been simulated both for the petal 
and for the entire lens. We expect to present these results in the next SPIE plenary conference in San Diego, 2013.

\section*{Acknowledgments}     
The LAUE project is the result of a big effort of many people and institutions. 
We wish to thank all of them. This work has been supported 
by the Agenzia Spaziale Italiana (ASI) through the project ''LAUE - $Una$ $Lente$ $per$ $i$ $raggi$ $Gamma$" under contract I/068/09/0.
Vineeth Valsan and Vincenzo Liccardo are supported by the Erasmus Mundus Joint Doctorate Program by Grant Number  
2010-1816 from the EACEA of the European Commission.

\bibliographystyle{spiebib}
\bibliography{SPIEref}

\end{document}